\documentstyle[12pt]{article}

\begin{document}

\begin{flushright}
IMSc/98/06/34 \\
hep-th/9806225
\end{flushright} 

\vspace{2mm}

\vspace{2ex}

\begin{center}
{\large \bf  Matrix Theory Description of Schwarzschild \\
             Black Holes in the Regime N $\gg$ S} \\ 

\vspace{6mm}
{\large  S. Kalyana Rama \footnote{On leave of absence from \\ 
Mehta Research Institute, Chhatnag Road, Jhusi, 
Allahabad 211 019, India.}}
\vspace{3mm}

Institute of Mathematical Sciences, C.I.T. Campus, \\ 
Taramani, Chennai 600 113, India \\ 
\vspace{1ex}
email: krama@imsc.ernet.in \\ 
\end{center}

\vspace{4mm}

\begin{quote}
ABSTRACT.  We study the description of Schwarzschild black
holes, of entropy $S$, within matrix theory in the regime $N
\stackrel{>}{_\sim} S \gg 1$. We obtain the most general matrix
theory equation of state by requiring that black holes admit a
description within this theory. It has a recognisable form in
various cases. In some cases a $D$ dimensional black hole can
plausibly be thought of as a $\tilde{D} = D + 1$ dimensional
black hole, described by another auxiliary matrix theory, but in
its $\tilde{N} \sim S$ regime. We find what appears to be a
matrix theory generalisation to higher dynamical branes of the
normalisation of dynamical string tension, seen in other
contexts. We discuss a further possible generalisation of the
matrix theory equation of state. In a special case, it is
governed by $N^3$ dynamical degrees of freedom.
\end{quote}

\newpage

{\bf 1.} In the past few years a lot of progress has been made
in understanding the properties of charged black holes using the
physics of Dirichlet branes. The latest advance in this progress
is in the study of Schwarzschild black holes in the context of
matrix theory \cite{bfss}. In \cite{bfks1,ks} Banks, Fischler,
Klebanov, and Susskind (BFKS) have shown that the matrix theory
compactified on $p$ dimensional torus reproduces, upto numerical
factors, the correct mass vs entropy relation for the $D$ $(= 11
- p)$ dimensional Schwarzschild black holes. Many other
properties such as Hawking radiation, long range interaction,
etc. have also been obtained \cite{bfks2,bfk}. The Schwarzschild
black holes, thought of as a collection of matrix partons, have
also been studied in the context of SYM theory and Dirichlet
0-branes \cite{bfks2,hm,li,limartinec}. For some of the other
related studies, see \cite{others}.

Matrix theory can be thought of as the discretised light cone
quantisation (DLCQ) of M theory \cite{bfss}, with $N$ units of
quantised longitudinal momentum. To describe adequately the
properties of a black hole with entropy $S$, $N$ must be atleast
of order $S$. Since the matrix theory computations are difficult
for $N \gg S$, BFKS compute the matrix theory equation of state
(e.o.s) for $N \sim S$ and show that it reproduces, upto
numerical factors, the correct mass vs entropy relation for the
Schwarzschild black holes.

The BFKS regime $N \sim S$ turns out to be where a black hole to
black string transition \cite{myers} takes place \cite{hm,das}.
See also \cite{hp}. Essentially, for $N < S$ the black hole does
not fit into the longitudinal space. This is a black string
configuration.  For $N > S$ the black hole Lorentz contracts and
fits into the longitudinal space easily. This is a black hole
configuration. Thus, $N \sim S$ is the regime where the
transition takes place. When $N \gg S$, the black hole size
becomes much smaller. For a precise explanation, see \cite{hm}.

The study of $N \gg S$ is of obvious interest and it is likely
to yield valuable insights. However, in this regime, the matrix
theory computations are difficult, and not much is known from
the SYM theory side either. In this regime, BFKS take the entropy
$S$ to be independent of $N$, but now equal to the black hole
entropy and thus predict the matrix theory e.o.s. They also show
that it is consistent with the physics of the SYM theory
\cite{bfks1,ks}. In a different analysis, Li and Martinec model
the black hole as a collection of matrix partons
\cite{bfks2,hm,li}, and study the regime $N \gg S$ by treating
the Lorentz contracted black hole as a collection of $S$
interacting clusters of $\frac{N}{S}$ units each
\cite{limartinec}.

In section 2, we give a brief description of the results of
BFKS. 

In this paper, we study the regime $N \gg S$ by another method.
We mainly study if there are regimes $N^c \sim S$, $0 < c \le 1$
so that $N \stackrel{>}{_\sim} S \gg 1$, where the resulting
e.o.s is of a recognisable form - particularly, of that of a
$\gamma + 1$ dimensional gas, with $\gamma$ a positive
integer. Such regimes are very likely to have a field theoretic
description.

For $N \gg S$, the size of the $D$ dimensional black hole
becomes much smaller than the size of the longitudinal
space. Hence, plausibly, we can think of such black holes as a
$\tilde{D} = D + 1$ dimensional black holes and expect it to be
described by another auxiliary matrix theory, in its BFKS
regime. We find that this is indeed the case. In section 3, we
give explicit illustration of various cases. However, we are
unable to make more precise the physical sense in which a $D$
dimensional black hole, in the regime $N^c \sim S$, $0 < c < 1$,
is equivalent to $\tilde{D} = D + 1$ dimensional black hole of
an auxuliary theory, but in its BFKS regime.

It turns out that the BFKS results as well those in section 3
can all be derived systematically by imposing two requirements:
(I) we work within the matrix theory framework, and (II) the $D$
dimensional Schwarzschild black holes admit a description within
the matrix theory compactified on a $p$ $(= 11 - D)$ dimensional
torus. These two requirements lead to the most general
expression for the matrix theory e.o.s. In section 4, we derive
this expression and study various new regimes where the matrix
theory e.o.s is of a recognisable form.

In the course of this study, we find that the tension of the
dynamical $\delta$-branes ($\delta$ is a positive integer),
defined in a particular natural sense, gets normalised as $N^{-
\frac{1 + \delta}{\delta}}$. This appears to be a matrix theory
generalisation to higher dynamical branes of a phenomenon seen
in other contexts where dynamical string tension is normalised  
as $\frac{1}{N^2}$ \cite{gkp,hk}.  

In section 5, we discuss a further possible generalisation of
the matrix theory e.o.s derived in the earlier section. This
generalisation goes beyond the realm of DLCQ framework. We find
one case which appears to be special and where the e.o.s in the
regime $N \gg S$ is governed by $N^3$ dynamical degrees of
freedom.

Although we believe that the present results are interesting,
we also admit to one crucial shortcoming: we are unable to
understand these results from the SYM theory side and, hence, are
unable to present a simple physical picture, if exists, that
underlies these results. In section 6 we conclude with a summary
and a critical remark.

{\bf 2.} In this section we present the relevent details of the
matrix theory description of $D$ dimensional Schwarzschild black
holes, as given in \cite{bfks1,ks,bfks2}. Thereby, we will set
our notation also.

Matrix theory can be thought of as the discretised light cone
quantisation (DLCQ) of M theory \cite{bfss}. Let $R$ be the
radius of the light like circle. The quantised longitudinal
momentum is then given by
\begin{equation}\label{p-}
P_- = \frac{N}{R} 
\end{equation}
where $N$ is an integer. The matrix theory Hamiltonian is
identified with the DLCQ energy according to
\begin{equation}\label{mx}
E = \frac{M^2}{P_-} = \frac{M^2 R}{N} \; . 
\end{equation}

To describe the $D$ dimensional Schwarzschild black holes, the
matrix theory is compactified on a $p$ $(= 11 - D)$ dimensional
torus. For simplicity, the torus is taken to be ``square'' with
all circumferences equal to $L$.  According to matrix theory
conjecture, the sector of the theory with given $N$ is described
exactly by the $p + 1$ dimensional U(N) SYM theory with 16 real
supercharges. The SYM theory lives on a dual square torus of
volume $V$ given by
\begin{equation}\label{vp}
V = \left( \frac{l^3_{11}}{R L} \right)^p  
\end{equation}
where $l_{11}$ is the eleven dimensional Planck length. 
The SYM coupling constant is given by 
\begin{equation}\label{g2}
g^2 = \frac{V R^3}{l_{11}^6} \; . 
\end{equation}
The effective SYM coupling constant for large $N$ is given by
\begin{equation}\label{e2}
e^2 = g^2 N \; . 
\end{equation}

In the $D$ dimensional non compact space time, the Newton's
constant $G_D$ and the Planck length $l_D$ are given by
\begin{equation}\label{ld}
G_D = l_D^{9 - p} = \frac{l_{11}^9}{L^p} 
\end{equation}
where we have used $D + p = 11$. The entropy $S$ of the $D$
dimensional Schwarzschild black hole of mass $M$ is then given,
upto numerical factors, by
\begin{equation}\label{sbh}
S \sim (l_D M)^{\frac{9 - p}{8 - p}} \; . 
\end{equation} 

To describe adequately the properties of a black hole with
entropy $S$, $N$ must be atleast of order $S$ as explained in
detail in \cite{bfks1}. However, since the matrix theory
computations are difficult for $N \gg S$, Banks et al (BFKS)
take $N \sim S$ \cite{bfks1,ks,bfks2}. Their strategy is to
first obtain the matrix model equation of state (e.o.s) $E = E(N,
S)$. For $N \sim S$, the matrix model Hamiltonian (\ref{mx})
gives the relation
\begin{equation}\label{bfks}
M^2 \sim \frac{S}{R} E(S, S) 
\end{equation}
between mass and entropy. This is then to be compared with the
mass vs entropy relation of the black hole, as explained in
\cite{bfks1}.

The matrix model e.o.s is obtained by its
conjectured relation to SYM theory. For $p = 3$, the conformal
invariance of the theory determines the e.o.s. For
other values of $p$, its description as $N$ coincident Dirichlet
$p$-branes determines the e.o.s \cite{kt2}. The
e.o.s thus obtained can be written as
\begin{equation}\label{eos}
S = \left( (N^2 V)^{5 - p} \; e^{2 (p - 3)} 
E^{9 - p} \right)^{\frac{1}{2 (7 - p)}} \; .
\end{equation}
Applying BFKS strategy now, namely taking $N \sim S$, equation
(\ref{eos}) can be seen to reproduce, upto numerical factors of
order unity, the correct mass vs entropy relation for the 
$D$ dimensional Schwarzschild black hole, $4 \le D \le 11$.

Note that (\ref{eos}) can also be written as 
\begin{equation}\label{n2sigmae}
S = \left( N^2 (\sigma E)^\gamma 
\right)^{\frac{1}{1 + \gamma}} 
\end{equation} 
where $\gamma = \frac{9 - p}{5 - p}$ and the parameter $\sigma$
has dimensions of length and is given by a specific combination
of the two dimensionful parameters $V$ and $e^2$
\begin{equation}\label{sigma1}
\sigma = V^{\frac{5 - p}{9 - p}} \;   
e^{\frac{2 (p - 3)}{9 - p}} \; . 
\end{equation} 
The significance of $\sigma$ will become clear later.  Thus, for
the $D = 7, 8, 10$ dimensional black hole, i.e. for $p = 4, 3,
1$, the e.o.s (\ref{eos}) is that of a $\gamma + 1$ dimensional
gas, $\gamma = 5, 3, 2$ respectively, with $N^2$ degrees of
freedom.  See \cite{bfks1,ks} for details.

The choice of the regime $N \sim S$ means the following
\cite{bfks1}. Let $R_s$ be the Schwarzschild radius of the black
hole in its rest frame. For $R_s > R$, the black hole does not
fit into the longitudinal space. When it is boosted
longitudinally such that its momentum is $P_- = \frac{N}{R}$,
its radius Lorentz contracts to
\begin{equation}\label{rs}
r_s = \frac{M}{P_-} R_s = \frac{M R}{N} R_s \; .
\end{equation}
In the regime $N \sim S \sim M R_s$, the Lorentz contracted
radius $r_s \sim R$ and thus the black hole just fits inside the
longitudinal space. A precise explanation is given in
\cite{hm}. The regime $N \sim S$ also turns out to be where the
black hole to black string transition \cite{myers} takes place
\cite{hm,das}. See also \cite{hp}.

When $N \gg S$, the entropy must be a constant, independent of
$N$. Taking the constant $S$ to be the black hole entropy $\sim
M R_s$, and using equation (\ref{mx}) and the thermodynamic
relation $d E = T d S$, one obtains the e.o.s:
\begin{equation}\label{eosnggs}
S = \left( N^{2 (6 - p)} V^{5 - p} \; e^{2 (p - 3)} 
E^{9 - p} \right)^{\frac{1}{2 (8 - p)}} \; .
\end{equation}
At $N \sim S$ this e.o.s agrees with that given in (\ref{eos}).
As this equation is derived using black hole physics alone, it
can also be thought of as an implication for the SYM e.o.s in
the regime $N \gg S$, and is, indeed, consistent with the
physics of SYM theory \cite{bfks1,ks}. Note that the length
scale appearing in (\ref{eosnggs}) is also given by
(\ref{sigma1}).

{\bf 3.} In this paper, we study the regime $N \gg S$ by another
method.  We mainly study if there are regimes $N^c \sim S$, $0 <
c \le 1$ so that $N \stackrel{>}{_\sim} S \gg 1$, where the
resulting e.o.s is of a recognisable form - particularly, of
that of a $\gamma + 1$ dimensional gas, with $\gamma$ a positive
integer. Such regimes are very likely to have a field theoretic
description.

In the regime $N \gg S$, the Lorentz contracted Schwarzschild
radius $r_s$ given by (\ref{rs}) becomes much smaller than $R$,
the size of the longitudinal space. Hence, the original
$D$ dimensional black hole can plausibly be thought of as a
$\tilde{D} = D + 1$ dimensional black hole. Such a black hole
may be expected to be described by another auxiliary matrix
theory, but now in its BFKS regime, say $\tilde{N} \sim S$ for a
suitable $\tilde{N}$.

Thus, specifically, in the regime
\begin{equation}\label{ncs}
N^c \sim S \; , \; \; \; 0 < c < 1 \; \; \; \; \; \; 
{\rm so} \; \; {\rm that} \; \; \; N \gg S \gg 1 \; , 
\end{equation}
we may expect the e.o.s of a $D$ dimensional black hole, for
suitable values of $D$, to be that of a $\gamma + 1$ dimensional
gas, where the value of $\gamma$ is the one appropriate for
$\tilde{D} = D + 1$ dimensional black hole, but in the BFKS
regime of the auxiliary matrix theory. We find that this is
indeed the case and now give explicit illustration. Since the
gas-like description is possible in the BFKS regime $\tilde{N}
\sim S$ of the auxiliary theory for $\tilde{D} = D + 1 = 7, 8,
10$ only, we study below the regime $N^c \sim S$, $0 < c < 1$
for the cases $D = 6, 7, 9$ only.

{\bf 3 a.}  Let $D = 6$. The e.o.s in the regime (\ref{ncs})
must be of the form
\[
S \sim E^{\frac{5}{6}} 
\]
which follows because $\gamma = 5$ is the appropriate value for
the $\tilde{D} = 7$ dimensional blackhole in the BFKS regime of
the auxiliary theory. Throughout this section, we write
explicitly the $N$ and $E$, equivalently $M$, dependence only.
Since the only relevent length scale in the e.o.s appears to be
the one given by (\ref{sigma1}), we take the e.o.s to be of the
form given in (\ref{n2sigmae}). Thus, we get
\begin{equation}\label{d6}
S \sim \left( N^{2 + \frac{5}{2}} E^5 
\right)^{\frac{1}{6}} \; . 
\end{equation}
The fraction $\frac{5}{2}$ arises from the $N$ dependence in
$\sigma$ in (\ref{n2sigmae}) where $p = 11 - 6 = 5$ and $\gamma
= 5$. Upon using (\ref{mx}) to write $E$ in terms of $M$ and
$N$ and then using (\ref{ncs}) to write $N$ in terms of $S$, we
get
\[
S^{6 + \frac{1}{2 c}} \sim M^{10} \; .
\]
For $c = \frac{1}{3}$, this leads to 
\[
S \sim M^{\frac{4}{3}}
\]
which is the correct mass vs entropy relation for the $6$
dimensional \\ 
Schwarzschild black hole, but now in the regime $N
\sim S^3$. Moreover, in the regime $N \sim S^3$, the e.o.s
(\ref{eosnggs}), which was obtained for $N \gg S$ with $S$ taken
to be constant, agrees with the above e.o.s (\ref{d6}). 

The Hawking temperature is also correctly reproduced after
deboosting the temperature $T_c$, corresponding to the regime $N
\sim S^3$, to the rest frame of the black hole. In the regime $N
\sim S^3$ the temperature $T_c$, obtained from (\ref{d6}) along
with the thermodynamic relation $d E = T d S$, is given by
\[
S \sim N^{2 + \frac{5}{2}} T_c^5 
\; \; \; \longrightarrow \; \; \;
T_c \sim N^{- \frac{5}{6}} \; . 
\]
By deboosting to the black hole rest frame we get the Hawking
temperature to be 
\[
T_H = \frac{N}{M} T_c \sim M^{- \frac{1}{3}}
\]
where we have used $N \sim S^3 \sim M^4$, valid in the regime
under consideration. The Schwarzschild radius in this regime is
given by 
\[
S \sim M R_s \; \; \; \longrightarrow \; \; \;
R_s \sim M^{\frac{1}{3}} 
\]
and, hence, we have $T_H \sim \frac{1}{R_s}$, the correct
scaling relation for Schwarzschild black hole.

{\bf 3 b.} Let $D = 7$. The e.o.s in the regime
(\ref{ncs}) must be of the form
\[
S \sim E^{\frac{3}{4}} 
\]
which follows because $\gamma = 3$ is the appropriate value for
the $\tilde{D} = 8$ dimensional blackhole in the BFKS regime of
the auxiliary theory. As above, taking the e.o.s to be of the
form given in (\ref{n2sigmae}), we get
\begin{equation}\label{d7}
S \sim \left( N^{2 + \frac{3}{5}} E^3 
\right)^{\frac{1}{4}} \; . 
\end{equation}
The fraction $\frac{3}{5}$ arises from the $N$ dependence in
$\sigma$ in (\ref{n2sigmae}) where $p = 11 - 7 = 4$ and $\gamma
= 3$. Upon using (\ref{mx}) to write $E$ in terms of $M$ and $N$
and then using (\ref{ncs}) to write $N$ in terms of $S$, we get
\[
S^{4 + \frac{2}{5 c}} \sim M^6 \; .
\]
For $c = \frac{1}{2}$, this leads to 
\[
S \sim M^{\frac{5}{4}}
\]
which is the correct mass vs entropy relation for the $7$
dimensional \\ 
Schwarzschild black hole, but now in the regime $N
\sim S^2$. Moreover, in the regime $N \sim S^2$, the e.o.s
(\ref{eosnggs}), which was obtained for $N \gg S$ with $S$ taken
to be constant, agrees with the above e.o.s (\ref{d7}). 

The Hawking temperature is also correctly reproduced after
deboosting the temperature $T_c$, corresponding to the regime $N
\sim S^2$, to the rest frame of the black hole. In the regime $N
\sim S^2$ the temperature $T_c$, obtained from (\ref{d7}) along
with the thermodynamic relation $d E = T d S$, is given by
\[
S \sim N^{2 + \frac{3}{5}} T_c^3 
\; \; \; \longrightarrow \; \; \;
T_c \sim N^{- \frac{7}{10}} \; . 
\]
By deboosting to the black hole rest frame we get the Hawking
temperature to be 
\[
T_H = \frac{N}{M} T_c \sim M^{- \frac{1}{4}}
\]
where we have used $N \sim S^2 \sim M^{\frac{5}{2}}$, valid in
the regime under consideration. The Schwarzschild radius in this
regime is given by
\[
S \sim M R_s \; \; \; \longrightarrow \; \; \;
R_s \sim M^{\frac{1}{4}} 
\]
and, hence, we have $T_H \sim \frac{1}{R_s}$, the correct
scaling relation for Schwarzschild black hole.

{\bf 3 c.} Let $D = 9$. The e.o.s in the regime
(\ref{ncs}) must be of the form
\[
S \sim E^{\frac{2}{3}} 
\]
which follows because $\gamma = 2$ is the appropriate value for
the $\tilde{D} = 10$ dimensional blackhole in the BFKS regime of
the auxiliary theory. As above, taking the e.o.s to be of the
form given in (\ref{n2sigmae}), we get
\begin{equation}\label{d9}
S \sim \left( N^{2 - \frac{2}{7}} E^2 
\right)^{\frac{1}{3}} \; .
\end{equation}
The fraction $- \frac{2}{7}$ arises from the $N$ dependence in
$\sigma$ in (\ref{n2sigmae}) where $p = 11 - 9 = 2$ and $\gamma
= 2$. Upon using (\ref{mx}) to write $E$ in terms of $M$ and
$N$ and then using (\ref{ncs}) to write $N$ in terms of $S$, we
get
\[
S^{3 + \frac{2}{7 c}} \sim M^4 \; .
\]
For $c = \frac{2}{3}$, this leads to 
\[
S \sim M^{\frac{7}{6}}
\]
which is the correct mass vs entropy relation for the $9$
dimensional \\ 
Schwarzschild black hole, but now in the regime $N
\sim S^{\frac{3}{2}}$. Moreover, in the regime $N \sim
S^{\frac{3}{2}}$, the e.o.s (\ref{eosnggs}), which was obtained
for $N \gg S$ with $S$ taken to be constant, agrees with the
above e.o.s (\ref{d9}). 

The Hawking temperature is also correctly reproduced after
deboosting the temperature $T_c$, corresponding to the regime $N
\sim S^{\frac{3}{2}}$, to the rest frame of the black hole. In
the regime $N \sim S^{\frac{3}{2}}$ the temperature $T_c$,
obtained from (\ref{d9}) along with the thermodynamic relation
$d E = T d S$, is given by
\[
S \sim N^{2 - \frac{2}{7}} T_c^2 
\; \; \; \longrightarrow \; \; \;
T_c \sim N^{- \frac{11}{21}} \; . 
\]
By deboosting to the black hole rest frame we get the Hawking
temperature to be 
\[
T_H = \frac{N}{M} T_c \sim M^{- \frac{1}{6}}
\]
where we have used $N \sim S^{\frac{3}{2}} \sim
M^{\frac{7}{4}}$, valid in the regime under consideration. The
Schwarzschild radius in this regime is given by
\[
S \sim M R_s \; \; \; \longrightarrow \; \; \;
R_s \sim M^{\frac{1}{6}} 
\]
and, hence, we have $T_H \sim \frac{1}{R_s}$, the correct
scaling relation for Schwarzschild black hole.

Thus, we have seen in the above cases that the $D$ dimensional
black hole in the regime $N^c \sim S$, $0 < c < 1$ can plausibly
be thought of as a $\tilde{D} = D + 1$ dimensional black hole
described by an auxiliary matrix theory, but in its BFKS regime
$\tilde{N} \sim S$. 

Clearly, the auxiliary matrix theory is to be compactified on a
$\tilde{p} = p - 1$ dimensional torus. The parameters of the
auxiliary matrix theory - the quanta, $\tilde{N}$, of the
longitudinal momentum, the radius, $\tilde{R}$, of the
light-like circle, the volume, $\tilde{V}$, of the dual
$\tilde{p}$ dimensional torus, and the effective coupling
constant, $\tilde{e}^2$, of the $\tilde{p} + 1$ dimensional SYM
theory - can all be determined explicitly by noting that the
physical parameters must be the same. Namely, the entropy $S$,
the mass $M$, and the longitudinal momentum $P_-$ of the black
hole must be the same in both the original and the auxiliary
matrix theories. We thus get
\begin{eqnarray}
\tilde{N} & = & N^c \nonumber \\ 
\tilde{R} & = & N^{c - 1} R \nonumber \\ 
\tilde{V} & = & N^{\frac{2 (p - 2)}{8 - p}} 
V^{\frac{7 - p}{9 - p}} e^{\frac{6}{9 - p}} \nonumber \\
\tilde{e}^2 & = & N^{\frac{2 (p - 6)}{8 - p}} 
V^{- \frac{2}{9 - p}} e^{\frac{2 (12 - p)}{9 - p}} 
\label{aux} \; .
\end{eqnarray}

However, we are unable to make more precise the physical sense
in which a $D$ dimensional black hole, in the regime $N^c \sim
S$, $0 < c < 1$, is equivalent to $\tilde{D} = D + 1$
dimensional black hole of an auxuliary theory, but in its BFKS
regime; equivalently, the physical sense in which the $p + 1$
dimensional SYM theory in the regime $N^c \sim S$, $0 < c < 1$,
is equivalent to an auxiliary $\tilde{p} + 1$ ($= (p - 1) + 1$)
dimensional SYM theory, but in its BFKS regime. The main problem
is that we lack a clear understanding of the e.o.s from the SYM
theory side in the regime $N^c \sim S$, $0 < c < 1$. The
relations in (\ref{aux}) may, perhaps, be useful in this regard.

{\bf 4.} The BFKS results as well as those in section 3 can all
be derived systematically by imposing two requirements: (I) we
work within the matrix theory framework, and (II) the $D$
dimensional Schwarzschild black holes admit a description within
the matrix theory compactified on a $p$ $(= 11 - D)$ dimensional
torus (taken, for simplicity, to be ``square'' with all
circumferences equal to $L$). These two requirements lead to the
most general expression for the matrix theory e.o.s. One may
then study the description of black holes in the BFKS regime $N
\sim S$ or in the generalisd regime $N^c \sim S$, $0 < c \le
1$. That is, in the regime $N \stackrel{>}{_\sim} S \gg 1$,
where the matrix theory has sufficient degrees of freedom to
describe the black hole with entropy $S$ \cite{bfks1}.

Through the matrix theory conjecture, the requirement II is
equivalent to the requirement that the $D$ dimensional
Schwarzschild black holes admit a description within the $p + 1$
dimensional U(N) SYM theory with 16 real supercharges.
Consequently, the resulting most general e.o.s can be viewed as
the most general e.o.s for the SYM theory, following from the
requirement that the $D$ dimensional Schwarzschild black holes
admit a description within this theory. 

The matrix theory Hamiltonian is given by (\ref{mx}), reproduced
below:
\[
E = \frac{M^2 R}{N} \; . 
\]
The matrix theory conjecture that the sector of the theory with
given $N$ is described exactly by the $p + 1$ dimensional U(N)
SYM theory with 16 real supercharges implies that the entropy
$S$ must be expressible in terms of the SYM parameters
only. Namely, in terms of $N$, $V$ the volume of the dual torus
given in (\ref{vp}), and $g^2$ (or equivalently $e^2$) given in
(\ref{g2}) (or (\ref{e2})), only. Thus, most generally the
entropy $S$ can be written as
\begin{equation}\label{sym}
S = \left( N^\alpha V^a e^{2 b} E^\gamma 
\right)^{\frac{1}{1 + \gamma}} 
\end{equation}
for some constants $\alpha, a, b$, and $\gamma$. Within matrix
theory, the $N$ dependence can not be determined. But the value
$\alpha = 2$, which implies that the dynamics is governed by
$N^2$ degrees of freedom, is natural in the context of U(N) SYM
theory. The parameter $\gamma$ is not necessarily an integer,
nor even positive. Since $S$ is dimensionless, we have
\begin{equation}\label{ap}
a p + b (p - 3) = \gamma 
\end{equation} 
since $V$ and $e^2$ have length dimensions $p$ and $(p - 3)$
respectively. Equations (\ref{sym}) and (\ref{ap}) are thus the
consequences of the requirement (I).

Now, the requirement (II) that the $D$ dimensional Schwarzschild
black hole admit a description in matrix theory implies that the
entropy $S$ must be expressible in terms of $N$, the $D$
dimensional Planck length $l_D$ (\ref{ld}), and the mass $M$ (or
equivalently $\frac{E}{R}$) only. Thus, most generally the
entropy $S$ can be written, noting that it is dimensionless, as
\begin{equation}\label{sg}
S = N^{A_1} \left( \frac{l_D^2 E}{R} \right)^{A_2} 
\end{equation}
for some constants $A_1$ and $A_2$. 

It follows, upon using equations (\ref{vp})-(\ref{ld}), that the
above expression is of the form (\ref{sym}) iff $a$ and $b$
satisfy the relation
\begin{equation}\label{bbya}
\frac{b}{a} = \frac{p - 3}{5 - p} \; . 
\end{equation}
Solving equations (\ref{ap}) and (\ref{bbya}) for $a$ and $b$
gives 
\begin{equation}\label{ab}
\frac{a}{\gamma} = \frac{5 - p}{9 - p} 
\; \; , \; \; \; \; \; \; 
\frac{b}{\gamma} = \frac{p - 3}{9 - p} \; .
\end{equation}

This implies that the scale which dictates the dynamics of
matrix theory is a length scale $\sigma$, or equivalently
$\sigma_0$, which is a specific combination of the dimensionful
SYM parameters $V$ and $e^2$, or equivalently $g^2$, and is given
by
\begin{equation}\label{sigma}
\sigma = V^{\frac{5 - p}{9 - p}} \;   
e^{\frac{2 (p - 3)}{9 - p}} \; , 
\end{equation} 
or equivalently by 
\begin{equation}\label{sigma0}
\sigma_0 = V^{\frac{5 - p}{9 - p}} \;   
g^{\frac{2 (p - 3)}{9 - p}} \; . 
\end{equation} 
The relevence of this scale to black hole physics can also be
glimpsed by noticing that it is related to the $D$ dimensional
Planck length $l_D$ by 
\begin{equation}\label{ld2r}
\sigma_0 = \frac{l_D^2}{R} \; . 
\end{equation}
Indeed, precisely this length scale enters in both the e.o.s
(\ref{eos}) and (\ref{eosnggs}) in the BFKS analysis in the
regimes $N \sim S$ and $N \gg S$ respectively.

Equating the two expressions (\ref{sym}) and (\ref{sg}) now
gives 
\begin{equation}\label{a1a2}  
A_1 = \frac{\alpha + b}{1 + \gamma} 
\; \; , \; \; \; \; \; \; 
A_2 = \frac{\gamma}{1 + \gamma} \; . 
\end{equation}

Thus, the matrix theory e.o.s (\ref{sym}) can be expressed in
any of the following forms:
\begin{eqnarray}
S & = & \left( N^\alpha (\sigma E)^\gamma 
\right)^{\frac{1}{1 + \gamma}} \label{ssigma} \\
& = & \left( N^{\alpha + b} (\sigma_0 E)^\gamma 
\right)^{\frac{1}{1 + \gamma}} \label{ssigma0} \\
& = & N^{\frac{\alpha + b}{1 + \gamma}} 
\left( \frac{l_D^2 E}{R} 
\right)^{\frac{\gamma}{1 + \gamma}} \label{sld} \\
& = & N^{\frac{\alpha + b - \gamma}{1 + \gamma}} 
(l_D M)^{\frac{2 \gamma}{1 + \gamma}}  \label{sm}
\end{eqnarray}
where $\alpha$ and $\gamma$ are constants, $b$ is given in
(\ref{ab}) and, in obtaining (\ref{sm}), we have used equation
(\ref{mx}) which relates the matrix theory Hamiltonian and the
DLCQ energy. This is the most general form of the matrix theory
e.o.s subject only to the matrix theory conjecture and to the
requirement that Schwarzschild black holes admit a description
within this theory. This can also be viewed, through the matrix
theory conjecture, as the most general e.o.s for the $p + 1$
dimensional U(N) SYM theory with 16 real supercharges, subject
only to the requirement that Schwarzschild black holes admit a
description within this theory.

We now adapt the BFKS strategy and require that the matrix
theory e.o.s reproduce the black hole entropy in the regime $N^c
\sim S$, $0 < c \le 1$. The entropy of the $D$ $(= 11 - p)$
dimensional Schwarzschild black hole is given, upto numerical
factors, by
\begin{equation}\label{sbh*}
S \sim (l_D M)^{\frac{9 - p}{8 - p}} \; . 
\end{equation} 
Hence, it follows after a simple algebra that the matrix theory
e.o.s reproduces black hole entropy in the regime $N^c \sim S$
iff $c$ and $\gamma$ satisfy the relation
\begin{equation}\label{cgamma}
c = \frac{2 \gamma (6 - p) - \alpha (9 - p)}{\gamma (7 - p) 
- (9 - p)} \; .
\end{equation}
For $c = 1$, equation (\ref{cgamma}) indeed yields the BFKS
result, namely $\gamma = \frac{9 - p}{5 - p}$.

We first describe two general results, which we have seen in
various cases previously. If
\begin{equation}\label{gamma*}
\gamma = \frac{9 - p}{7 - p} \equiv \gamma_* 
\end{equation} 
and, furthermore, if 
\begin{equation}\label{alpha*}
\alpha = \gamma - b = \frac{2 (6 - p)}{7 - p}
\end{equation} 
where we have used (\ref{ab}) and (\ref{gamma*}), then the
matrix theory e.o.s (\ref{sm}) is independent of $N$ and always
reproduces the black hole e.o.s. Expressed in terms of $E$, for
example in the form given in (\ref{ssigma0}), the matrix theory
e.o.s becomes
\begin{equation}\label{snggs}
S = (N \sigma_0 E)^{\frac{\gamma_*}{1 + \gamma_*}} \; . 
\end{equation}
As can be checked easily, this equation is same as equation
(\ref{eosnggs}) derived a la BFKS \cite{bfks1,ks}.

We have seen in various cases previously that the e.o.s
(\ref{snggs}), equivalently (\ref{eosnggs}), agrees with the
e.o.s obtained in the regimes $N^c \sim S$, $c \le 1$. From the
above derivation, it is now clear that this result is valid in
general. That is, valid for {\em any $c$ and any $D$}. This can
be checked explicitly in other ways also.

Another result, which we have seen in various cases previously,
is also valid in general {\em i.e.} valid for {\em any $c$ and
any $D$}: That deboosting the critical temperature $T_c$ in the
regime $N^c \sim S$ to the black hole rest frame gives the
Hawking temperature $T_H$ with the correct scaling. This can be
seen as follows. The e.o.s can be written generally
as
\begin{equation}\label{sae}
S = \left( {\cal A} E^\gamma 
\right)^{\frac{1}{1 + \gamma}} 
\end{equation} 
where ${\cal A}$ is a constant independent of $E$. Together with
the thermodynamic relation $d E = T d S$, this gives 
\begin{equation}\label{sat}
S = {\cal A} T^\gamma \; . 
\end{equation} 
In the regime $N^c \sim S \equiv S_c$, the critical temperature
$T_c$ is given by
\[
T_c \sim \left( \frac{S_c}{{\cal A}} \right)^{\frac{1}{\gamma}} 
\sim \frac{E}{S_c} 
\]
where the second step follows from (\ref{sae}). By deboosting to
the black hole rest frame we get the Hawking temperature to be
\[
T_H = \frac{N}{R M} T_c \sim 
\frac{N}{R M} \; \frac{E}{S_c} \; .
\]
Upon using (\ref{mx}) to write $E$ in terms of $M$, and noting
that $S_c \sim M R_s$ we get the Hawking temperature to be
\[
T_H \sim \frac{M}{S_c} \sim \frac{1}{R_s} 
\]
which is the correct scaling relation for Schwarzschild black
hole. Thus, this result is also true for {\em any $c$ and any
$D$}.

Now we return to equation (\ref{cgamma}), which ensures that the
matrix theory e.o.s reproduces the black hole entropy in the
regime $N^c \sim S$, $0 < c \le 1$. We study the regimes where
the e.o.s is that of a $\gamma + 1$ dimensional gas, with
$\gamma$ an integer. We also take $\alpha = 2$, the value which
is natural within the context of U(N) SYM theory.

\vspace{2ex}

\noindent{\bf (i)} In the BFKS regime, $c = 1$ and (\ref{cgamma}) 
gives $\gamma = \frac{9 - p}{5 - p} = \frac{D - 2}{D - 6}$, the
result obtained in \cite{bfks1,ks}.

\vspace{2ex}

\noindent{\bf (ii)} In the regime $N^c \sim S$, $0 < c < 1$,
let $\gamma = \frac{10 - p}{6 - p}$, as appropriate for the
$\tilde{D} = D + 1$ dimensional black hole in the BFKS regime of
the auxiliary theory. Equation (\ref{cgamma}) then gives $c =
\frac{6 - p}{8 - p}$. For $D = 6, 7, 9$, equivalently $p = 5, 4,
2$, we get $c = \frac{1}{3}, \frac{1}{2}, \frac{2}{3}$
respectively, which is the result obtained in section 3.

\vspace{2ex}

\noindent{\bf (iii)} One can now search for other regimes, {\em
i.e.} for values of $c$, if any, where the e.o.s is that of a
$\gamma + 1$ dimensional gas, with $\gamma$ an integer.

\vspace{2ex}

\noindent{\bf (a)} For $p \le 3$, there are no new regimes
other than the ones given in \cite{bfks1,ks} and in section 3.

\vspace{2ex}

\noindent{\bf (b)} For $p = 4$, besides the regimes given in
\cite{bfks1,ks} and in section 3, there is another regime with
$c = \frac{6}{7}$ and $\gamma = 4$. That is, the e.o.s of the
$7$ dimensional black hole in the regime $N \sim
S^{\frac{7}{6}}$ is that of a $4 + 1$ dimensional gas.

\vspace{2ex}

\noindent{\bf (c)} For $p = 5$, besides the regimes given in
\cite{bfks1,ks} and in section 3, there is a series of regimes
for every integer $\gamma \ge 6$ with $c = \frac{\gamma -
4}{\gamma -2} \le 1$. That is, the e.o.s of the $6$ dimensional
black hole in the regime $N \sim S^{\frac{1}{c}}$ is that of a
$\gamma + 1$ dimensional gas, for a series of values of $c$ and
$\gamma$ given as above.

In the BFKS regime, $c = 1$, $\gamma = \infty$ formally, 
indicating infinite specific heat. The e.o.s
(\ref{ssigma}) becomes
\[
S \sim e E 
\] 
which is the e.o.s of a dynamical string with tension $\sim
\frac{1}{N}$ \cite{ks}.

\vspace{2ex}

\noindent{\bf (d)} We will consider $p = 7$, before proceeding
to $p = 6$. There are no new regimes except the $N \gg S$ regime
considered in \cite{bfks2}. In this regime, $\gamma = \gamma_* =
\infty$ formally, indicating infinite specific heat, and $\alpha
+ b = \gamma$. See equations (\ref{gamma*}) and
(\ref{alpha*}). The e.o.s (\ref{ssigma0}) becomes
\begin{equation}\label{delta1}
S \sim N \sigma_0 E 
\end{equation}
which is the e.o.s of a dynamical string with tension $\sim
\frac{1}{N^2 \sigma_0^2}$ \cite{bfks2,gkp,hk}. The string length
scale, $\lambda_1$, of this dynamical string is given by
\begin{equation}\label{l1}
\lambda_1 \sim N \sigma_0 \; . 
\end{equation} 
Note that the string length scale $\lambda_1$ scales linearly
with $N$.

\vspace{2ex}

\noindent{\bf (e)} For $p = 6$ in the BFKS regime, $c = 1$ and
equation (\ref{cgamma}) gives $\gamma = - 3$ formally,
indicating negative specific heat. The e.o.s (\ref{ssigma0})
becomes
\begin{equation}\label{delta3}
S = \left( N^{\frac{1}{3}} \sigma_0 E 
\right)^{\frac{3}{2}} \; . 
\end{equation}
The e.o.s of a dynamical $\delta$-brane ($\delta$ is
a positive integer) is of the form \cite{kt}
\begin{equation}\label{delta}
S \sim E^{\frac{2 \delta}{1 + \delta}} \equiv 
(\lambda_\delta E)^{\frac{2 \delta}{1 + \delta}} 
\end{equation}
where $\lambda_\delta$ is a constant, of length dimension 1.
Hence, we see that e.o.s (\ref{delta3}) has the form
characteristic of that of dynamical 3-branes \cite{ks}. That the
dynamics here involves some $3$ (spatial) dimensional object is
not surprising in the light of the indications that the matrix
theory compactified on 6 dimensional torus, {\em i.e.} $p = 6$,
seems to involve gravitational degrees of freedom of $3 + 1$
dimensional nature \cite{ss}.

For dynamical $\delta$-branes, $\delta \ne 1$, there does not
appear to be a canonical way of reading out the brane tension,
$T_\delta$, from the e.o.s (\ref{delta}) alone. However, a
natural choice would be to identify $\lambda_\delta$, the
coefficient of $E$ in (\ref{delta}), as the dynamical brane
length scale, upto numerical factors. The tension $T_\delta$ is
then given by
\begin{equation}\label{tdelta}
T_\delta = (\lambda_\delta)^{- (1 + \delta)} \; .
\end{equation}
For $\delta$-branes, we can also define a brane volume scale: 
\begin{equation}\label{vdelta}
v_\delta = (\lambda_\delta)^\delta \; , 
\end{equation}
which can be thought of as a brane analog of the string length
scale.

Thus, with the natural choice of the brane parameters given
above, the coefficient of $E$ in (\ref{delta3}) can be
interpreted as $\lambda_3$, the dynamical 3-brane length
scale. The tension and the brane volume scale of the dynamical
3-brane are then given by
\begin{equation}\label{tv3}
T_3 \sim \frac{1}{N^{\frac{4}{3}} \sigma_0^4} \; , \; \; \; \;  
v_3 \sim N \sigma_0^3 \; . 
\end{equation} 
Note that the brane volume scale $v_3$ scales linearly with
$N$. 

There is also another regime $N^c \sim S$, $c < 1$, where the
e.o.s has a recognisable form. This regime is given by $c =
\frac{6}{7}$, {\em i.e.} the regime where $N \sim
S^{\frac{7}{6}}$. In this regime $\gamma = - 4$ formally,
indicating negative specific heat. The e.o.s (\ref{ssigma0})
becomes
\begin{equation}\label{delta2}
S = \left( N^{\frac{1}{2}} \sigma_0 E \right)^{\frac{4}{3}} 
\end{equation}
which has the form characteristic of that of dynamical 2-branes.
Note that the dynamical region (2-brane) now has one less
dimension than that (3-brane) in the BFKS regime. This is
reminiscent of the phenomenon we have seen in section 3 where
$D$ dimensional black hole in the regime $N^c \sim S$, $0 < c <
1$ can plausibly be thought of as described by a $\tilde{D} = D
+ 1$ dimensional black hole in the BFKS regime of the auxiliary
theory. The auxiliary theory is to be compactified on a
$\tilde{p} = p - 1$ dimesnional torus. The corresponding
auxiliary SYM theory lives on a $\tilde{p} + 1$ dimensional space
time, which indeed has one less dimension than that ($p + 1$
dimensional SYM theory) in the BFKS regime.

With the natural choice of the brane parameters given above, the
coefficient of $E$ in (\ref{delta2}) can be interpreted as
$\lambda_2$, the dynamical 2-brane length scale. The tension and
the brane volume scale of the dynamical 2-brane are then given
by
\begin{equation}\label{tv2}
T_2 \sim \frac{1}{N^{\frac{3}{2}} \sigma_0^3} \; , \; \; \; \; 
v_2 \sim N \sigma_0^2 \; . 
\end{equation} 
Note that the brane volume scale $v_2$ scales linearly with
$N$. 

In all the above cases where a dynamical $\delta$-brane ($\delta
= 1, 2, 3$, see equations (\ref{delta1}), (\ref{delta2}), and
(\ref{delta3})) appears, the brane tension $T_\delta$ is 
normalised w.r.t. $N$ according to
\begin{equation}\label{tn}
T_\delta \sim N^{- \frac{1 + \delta}{\delta}} \; .
\end{equation}
In the case of dynamical strings, $\delta = 1$ and the
normalisation becomes $T_1 \sim \frac{1}{N^2}$. For this case,
such normalisation has appeared in other contexts also. For
example in \cite{gkp} - where the dynamical open strings
attached to the $(N, 1)$ strings have an effective tension $\sim
\frac{1}{N^2}$. See \cite{hk} for another example. Equation
(\ref{tn}) appears to be a matrix theory generalisation of this
normalisation to higher dynamical branes. Also, very
interestingly, the brane-volume scale $v_\delta =
(\lambda_\delta)^\delta$, defined in analogy with the string
length scale, has a simple scaling in all the above cases:
\begin{equation}\label{vn}
v_\delta = (\lambda_\delta)^\delta \sim N \sigma_0^\delta \; . 
\end{equation}
Namely, the brane volume scale $v_\delta$ scales linearly with
$N$.

This is an intriguing result. It has a striking similarity to
the phenomenon seen in the case of Dirichlet branes, where
multiply wound single Dirichlet brane (with winding number $N$)
dominate over singly wound multiple branes ($N$ in number). The
effective volume of the multiply wound brane is $N$ times the
volume of the single brane. That is, the effective volume scales
linearly with $N$ \cite{dm}. However, here, it is the brane
volume scale $v_\delta$ - defined in (\ref{vdelta}) in analogy
with the string length scale - which scales linearly with
$N$. Clearly, $v_\delta$ is not the same as the volume of any
brane, atleast not in any obvious sense. We have no simple
explanation for this intriguing result.

{\bf 5.} The most general matrix theory e.o.s, given in
equivalent forms in (\ref{ssigma})-(\ref{sm}), is obtained by
requiring that the $D$ dimensional Schwarzschild black holes
admit a description within matrix theory compactified on a $p$
$(= 11 - D)$ torus. As such, it can be viewed, perhaps more
properly, as the most general e.o.s for the $p + 1$ dimensional
U(N) SYM theory with 16 real supercharges, subject only to the
requirement that the Schwarzschild black holes admit a
description within this theory. Equivalently, it can also be
thought of as the e.o.s for the dynamics of $N$ Dirichlet
$p$-branes, related to $N$ Dirichlet $0$-branes by T-dualities.

From this point of view, the general e.o.s is that given in
(\ref{sym}). Requiring that black holes admit a description in
this theory implies that (\ref{sym}) is expressible as in
(\ref{sg}) with various constants related by (\ref{ab}) and
(\ref{a1a2}). The resulting e.o.s can be written in any of the
equivalent forms (\ref{ssigma})-(\ref{sld}). Note, however, that
$R$ appears in the combination $\frac{E}{R}$ only.

Now, one further generalisation is possible. Equation (\ref{mx})
relating $E$ to $M$ and $N$ can be generalised to 
\begin{equation}\label{mxb}
E = \frac{M^2 R}{N^B} 
\end{equation}
where $B$ is a constant. That this is the only possible
generalisation follows from dimensional analysis and the fact
that $R$ appears in the combination $\frac{E}{R}$ only.
However, we do not know the physical significance, if any, of
the parameter $B$ when $\ne 1$. It may be that there is a
Seiberg-Sen type scaling \cite{ss} at work here, but which
involves $N$ also. 

The e.o.s now becomes 
\begin{equation}\label{smb}
S = N^{\frac{\alpha + b - \gamma B}{1 + \gamma}} 
(l_D M)^{\frac{2 \gamma}{1 + \gamma}}  \; .
\end{equation}
Requiring that it reproduce the black hole entropy in the regime
$N^c \sim S$, $0 < c \le 1$ then implies the relation: 
\begin{equation}\label{cgammab}
c = \frac{\gamma (3 (1 + 3 B) - p (1 + B)) - \alpha (9 - p)}
{\gamma (7 - p) - (9 - p)} \; .
\end{equation}
When $B = 1$ these two equations give (\ref{sm}) and
(\ref{cgamma}) respectively. 

As before, if $\gamma = \gamma_*$ given in (\ref{gamma*}) and,
furthermore, if
\begin{equation}\label{alpha*b}
\alpha = \gamma B - b = 
\frac{3 (1 + 3 B) - p (1 + B)}{7 - p} \; .
\end{equation} 
where we have used (\ref{ab}) and (\ref{gamma*}), then the
matrix theory e.o.s (\ref{smb}) is independent of $N$ and always
reproduces the black hole e.o.s.  The e.o.s is applicable in the
regime $N \gg S$, for reasons similar to those given in
\cite{bfks1,ks}.

We now look for special value(s) of $B$, if any, other than
$1$. Note that $\alpha$, given in equation (\ref{alpha*b}),
depends on $p$ in general.  Equivalently, equation
(\ref{alpha*b}) can be thought of as straight lines, drawn in
the $(B, \alpha)$ plane, one for each value of $p$ with slope
$\frac{9 - p}{7 - p}$. It turns out that these lines all
intersect at a common point in the $(B, \alpha)$ plane. That
more than two lines intersect at one point is evidently non
trivial, and the common intersection point, given by
\begin{equation}\label{b2}
(B, \alpha) = (2, 3) \; , 
\end{equation}
may perhaps signify a special value for $B$ and $\alpha$.  For
these values of $(B, \alpha)$, the e.o.s (\ref{ssigma}) becomes
\begin{equation}
S = \left( N^3 (\sigma E)^\gamma \right)^{\frac{1}{1 + \gamma}} 
\end{equation}
which implies that the dynamics is governed by $N^3$ degrees of
freedom. Such $N^3$ degrees of freedom also appear in the
entropy of N coincident 5-branes \cite{kt2} where they are
expected to be related to the appearance of tensionless strings
\cite{sw}. However, it is far from clear whether this is related
to the $N^3$ degrees of freedom appearing in the above e.o.s,
applicable in the $N \gg S$ regime, when $(B, \alpha)$ take the
special value $(2, 3)$.

{\bf 6.} To summarise: In this paper, we have studied the
description of Schwarzschild black holes within matrix theory in
the regime $N \stackrel{>}{_\sim} S \gg 1$, more specifically in
the regime $N^c \sim S$, $0 < c \le 1$. We mainly studied the
regimes where the e.o.s is of a recognisable form. We found that
in some cases a $D$ dimensional black hole can plausibly be
thought of as a $D + 1$ dimensional black hole, described by
another auxiliary matrix theory, but in its BFKS regime.

We then derived systematically these and the BFKS results by
imposing two requirements: (I) we work within the matrix theory
framework, and (II) the $D$ dimensional Schwarzschild black
holes admit a description within the matrix theory.
Consequently, we obtained the most general expression for the
matrix theory e.o.s and studied various new regimes where the
matrix theory e.o.s is of a recognisable form. In the course of
this study, we found what appears to be a matrix theory
generalisation to higher dynamical branes of the normalisation
of dynamical string tension, seen in other contexts.

We also discussed a further possible generalisation of the
matrix theory e.o.s mentioned above and found, in a special
case, that the e.o.s is governed by $N^3$ dynamical degrees of
freedom.

Although we believe that the present results are interesting,
there is a crucial shortcoming. We are unable to explain these
results from the SYM theory side. The main problem is that we
lack a clear understanding of the e.o.s from the SYM theory side
in the regime $N^c \sim S$, $0 < c < 1$. Therefore, we are
unable to present a simple physical picture, if exists, that
underlies these results. It will indeed be highly satisfactory
to derive the present results from the SYM theory side or,
conversely, understand their implication to the physics of SYM
theory. We are presently studying these issues.

\vspace{2ex}

{\bf Acknowledgements:}

We thank B. Sathiapalan for discussions.


\vspace{2ex}

\begin{thebibliography}{999}
\bibitem{bfss}
T. Banks, W. Fischler, S. H. Shenker, and L. Susskind, 
Phys. Rev. {\bf D55} (1997) 5112, hep-th/9610043; 
L. Susskind, hep-th/9704080.
For a review of matrix theory, see 
T. Banks, Nucl. Phys. Proc. Suppl. {\bf 67} (1998) 180, 
hep-th/9710231; 
D. Bigatti, and L. Susskind, hep-th/9712072. 
\bibitem{bfks1}
T. Banks, W. Fischler, I. R. Klebanov, and L. Susskind, 
Phys. Rev. Lett. {\bf 80} (1998) 226, hep-th/9709091. 
\bibitem{ks}
I. R. Klebanov, and L. Susskind, 
Phys. Lett. {\bf B416} (1998) 62, hep-th/9709108. 
\bibitem{bfks2}
T. Banks, W. Fischler, I. R. Klebanov, and L. Susskind, 
Jl. High Energy Phys. 01 (1998) 008, hep-th/9711005.  
\bibitem{bfk}
T. Banks, W. Fischler, and I. R. Klebanov, 
Phys. Lett. {\bf B423} (1998) 54, hep-th/9712236. 
\bibitem{hm}
G. T. Horowitz, and E. J. Martinec, 
Phys. Rev. {\bf D57} (1998) 4935, hep-th/9710217. 
\bibitem{li}
M. Li, Jl. High Energy Phys. 01 (1998) 009, hep-th/9710226. 
\bibitem{limartinec}
M. Li, and E. J. Martinec, hep-th/9801070. 
\bibitem{others}
E. Halyo, hep-th/9709225; 
H. Liu, and A. A. Tseytlin, 
Jl. High Energy Phys. 01 (1998) 010, hep-th/9712063; 
F. Englert, and E. Rabinovici, hep-th/9801048; 
R. Argurio, F. Englert, and L. Houart, hep-th/9801053. 
\bibitem{myers}
R. C. Myers, Phys. Rev. {\bf D35} (1987) 455;  
R. Gregory and R. Laflamme, 
Phys. Rev. Lett. {\bf 70} (1993) 2837, hep-th/9301052. 
\bibitem{das}
S. R. Das, S. D. Mathur, S. Kalyana Rama, and P. Ramadevi, 
to appear in Nucl. Phys. {\bf B}, hep-th/9711003.
\bibitem{hp}
G. T. Horowitz, and J. Polchinski, 
Phys. Rev. {\bf D55} (1997) 6189, hep-th/9612146. 
\bibitem{gkp}
S. Gukov, I. R. Klebanov, and A. M. Polyakov, 
Phys. Lett. {\bf B423} (1998) 64, hep-th/9711112. 
\bibitem{hk}
A. Hanany, and I. R. Klebanov, 
Nucl. Phys. {\bf B482} (1996) 105, hep-th/9606136. 
\bibitem{kt2}
I. R. Klebanov, and A. A. Tseytlin, 
Nucl. Phys. {\bf B475} (1996) 164, hep-th/9604089. 
\bibitem{kt}
I. R. Klebanov, and A. A. Tseytlin, 
Nucl. Phys. {\bf B479} (1996) 319, hep-th/9607107. 
For earlier references, see 
S. Fubini, A. J. Hanson, and R. Jackiw, 
Phys. Rev. {\bf D7} (1973) 1732; 
J. Dethlefsen, H. B. Nielsen, and H. C. Tze, 
Phys. Lett. {\bf B48} (1974) 48;  
A. Strumia, and G. Venturi, 
Lett. Nuomo Cimento {\bf 13} (1975) 337; 
E. Alvarez, and T. Ortin, 
Mod. Phys. Lett. {\bf A7} (1992) 2889; 
B. Harms, and Y. Leblanc, 
Phys. Rev. {\bf D47} (1993) 2438; 
A. A. Bytsenko, K. Kirsten, and S. Zerbini, 
Phys. Lett. {\bf B304} (1993) 235. 
\bibitem{ss}
A. Sen, hep-th/9709220; N. Seiberg, 
Phys. Rev. Lett. {\bf 79} (1997) 3577, hep-th/9710009. 
\bibitem{dm}
S. R. Das, and S. D. Mathur, 
Phys. Lett. {\bf B375} (1996) 103, hep-th/9601152; 
J. M. Maldacena, and L. Susskind, 
Nucl. Phys. {\bf B475} (1996) 679, hep-th/9604042. 
\bibitem{sw} 
E. Witten, hep-th/9507121; A.Strominger, 
Phys. Lett. {\bf B383} (1996) 44, hep-th/9512059.

\end{thebibliography}
\end{document}